\documentclass[sigconf]{acmart}
\AtBeginDocument{%
  }

\renewcommand\footnotetextcopyrightpermission[1]{}

\begin{document}

\title[Shared Data Infrastructure in Mandated Collaboration]{Collaboration by Mandate: How Shared Data Infrastructure Shapes Coordination and Control in U.S. Homelessness Services}

\thanks{This paper was accepted to the Extended Abstracts of the 2026 CHI Conference on Human Factors in Computing Systems (CHI EA '26).}

\author{Lingwei Cheng}
\affiliation{%
  \institution{School of Public Administration, University of Central Florida}
  \city{Orlando}
  \state{Florida}
  \country{USA}
}
\email{lingwei.cheng@ucf.edu}

\author{Saerim Kim}
\affiliation{%
  \institution{Department of Public Service and Healthcare Administration, Suffolk University}
  \city{Boston}
  \state{Massachusetts}
  \country{USA}
}
\email{saerim.kim@suffolk.edu}

\author{Andrew Sullivan}
\affiliation{%
  \institution{School of Public Administration, University of Central Florida}
  \city{Orlando}
  \state{Florida}
  \country{USA}
}
\email{andrew.sullivan@ucf.edu}

\renewcommand{\shortauthors}{Cheng et al.}

\begin{abstract}
When governments mandate collaboration, shared data systems can serve both as tools for coordination and instruments of control. This study examines U.S. homelessness service networks, where Continuums of Care (CoCs) coordinate service providers through the federally mandated Homeless Management Information System (HMIS). With client consent, providers enter data into HMIS and access cross-provider service histories to support coordinated care. At the same time, HMIS embeds standards and governance rules that shape who can collect, access, interpret, and act on data, and thus who holds decision authority. Using qualitative interviews with six experts, we show that standardization can facilitate collaboration and shared learning. However, unequal resources, analytic capacity, and authority limit equitable participation and often shift some participants toward compliance-focused roles. We contribute to public-interest design research on civic data infrastructures by illustrating how mandated data sharing can simultaneously enable coordination and accountability while reproducing power asymmetries in data interpretation and decision-making.
\end{abstract}



\maketitle

\section{Introduction}
Mandated collaboration --- where an external authority compels organizations to coordinate through laws, regulations, and funding requirements --- is common in fragmented social policy domains such as homelessness, child welfare, and disaster response \cite{agranoff2003collaborative, imperial2005using, sullivan2024critical}. In these domains, service delivery relies heavily on local nonprofit providers, creating incentives for government to require coordination. In the U.S., these mandates often operate through data, as federal and state governments embed data infrastructures into the core of mandated collaboration and increasingly tie funding and performance metrics to data reporting \cite{mcnamara2012starting, moulton2012preserving}. In such settings, data become the currency of coordination, accountability, and control: data-driven collaboration can shape who receives services, how funding is allocated, how performance is assessed, and how collaborators are held accountable. 

In social service provision, the design of data systems can either empower nonprofit service providers as strategic collaborators or create a cycle of dis-empowerment through erosion of autonomy, data drift, and data fragmentation \cite{bopp2017disempowered}. Policymakers, in turn, risk assuming that data mandates inherently foster collaboration when, in fact, they may reinforce power asymmetries or embed inequities in service delivery. 

Prior work on data and algorithms in homelessness has examined how frontline workers collect data and make decisions \cite{karusala2019street, kuo2023frontlineworkerhomeless}, and the implications for value tradeoff \cite{showkat23}, data quality \cite{moon2025datafication}, predictive validity, fairness, and perceptions of algorithmic tools \cite{moon2024human, chelmis21}. We know less about how shared data infrastructures structure data work and decision authority across the service network.

We address this gap through a case study of the U.S. homelessness system, where the federally mandated Continuums of Care (CoCs) service networks coordinate providers using a mandated shared data infrastructure: the Homeless Management Information System (HMIS). HMIS standardizes client-level data collection and reporting across service providers, enabling shared access to clients’ service histories and supporting system-wide coordination and accountability. Yet the same standards, access rules, and interpretive practices can centralize authority and exacerbate inequities between government and nonprofit providers \cite{gazley2020nonprofit, gronbjerg2016scope}. We ask: (1) How does a mandated shared data infrastructure structure cross-organizational data practices across the data lifecycle (collection, storage, analysis, and use)? (2) How does this infrastructure shape participation and influence? Specifically, do shared data practices support distributed, collaborative decision-making or concentrate control? 

Drawing on six interviews with stakeholders who manage and use HMIS in the Central Florida region, we show that effective data-driven collaboration depends not only on shared infrastructure but also on how access, analytical capacity, and interpretive authority are distributed across participants. Standardization supports interoperability and accountability, but it can limit local customization and place greater interpretive responsibility on the lead agency and more resourced organizations. Collaboration is further shaped by trust and alignment of incentives across partners.

We contribute to CHI research on civic technologies and public interest data infrastructures in three ways. First, we provide an in depth qualitative study of a civic data infrastructure, HMIS, showing how it shapes collaboration, accountability, and power across government and nonprofit actors. Second, we examine not only how data is collected through shared infrastructure, but also how it is collectively analyzed and transformed into a community narrative, and how frictions such as shifts toward compliance focused roles can concentrate authority. Third, we situate homelessness related HCI research within its broader policy and organizational context, highlighting how data governance and data infrastructures jointly shape data production, interpretation, and fairness.

\section{Related Work}
Mandated collaboration occurs when a third party requires organizations to cooperate and enforces that cooperation \citep{sullivan2024critical}. Mandates facilitate coordination by reducing transaction costs, excessive competition, or low inter-organizational trust. At the same time, they reshape accountability when participants are now responsible not only for their own performance but also for collective outcomes, creating potential conflicts over workload and authority. 

Crucially, mandated collaboration often operates through data: reporting standards, access rules, and accountability routines define performance and compliance requirements. We use \textit{data-driven collaboration} to refer to cross-organizational coordination achieved through shared data practices and management. 

Data infrastructures shape the balance between collaborative problem-solving and hierarchical control in inter-organizational governance. In distributed systems, they enable negotiation and collective problem-solving, while centralized designs often concentrate interpretive authority and decision-making \cite{ansell2008collaborative, emerson2012integrative}. Prior work on data use in nonprofits shows that when external stakeholders, especially funders, define performance metrics, tracking, and reporting, they can undermine nonprofit autonomy, leading to data drift, fragmentation, and disempowerment \cite{bopp2017disempowered}. Similarly, \citet{kawakami24studyup} show that power in government AI adoption operates across multiple levels --- within agencies, across levels of government, and downward toward clients -- revealing a disconnect between frontline concerns and leadership’s valuation of AI, as well as federal guidance that overlooks local resource constraints and weakly aligns private-sector partners with public goals. Even when there is broad recognition of the value of data tools, perceived local misalignment drives demands for customization \cite{venkat25static}. When collaboration is weak, mandates may drain resources and intensify conflict \cite{breimo2017networking, cain2020challenge}. In short, the governance embedded in data infrastructures can determine whether mandates enable collective problem-solving or instead intensify resource drain, misalignment, and conflict.

The effectiveness of shared data infrastructure is shaped by resources, capacity, and organizational context. Service providers must balance direct service provision with data work \cite{mosley2021cross, mcnamara2012starting, gazley2020nonprofit}. Incomplete reporting, limited software functionality, cost barriers, and uneven technical expertise are major hurdles to effectively leverage shared data \cite{ruege2023qualitydatainfrastructure}. Building the social relationships and organizational capacity needed for collaborative data practice is also resource intensive \cite{casey2016interorganizational}.  When resources are scarce, participants may be unable to sustain meaningful data practices, leading to uneven participation and influence across the network \cite{bopp2017disempowered}. Lastly, funders often tie participation and reporting to financial incentives \cite{mcnamara2012starting, moulton2012preserving}, limiting flexibility by binding scarce resources to compliance \cite{bopp2017disempowered}.

Together, these findings suggest that while data infrastructures can support collective learning and coordination, they may simultaneously reinforce disparities in capacity, influence, and decision-making, particularly in mandated networks where reporting and funding are tightly linked.

\section{Background}

We investigate how mandated data sharing shape collaboration and power in U.S. homelessness services, where federally mandated Continuums of Care (CoCs) coordinate regional providers and act as the liaison to the Department of Housing and Urban Development (HUD). CoCs typically designate a lead agency to coordinate activities and to operate the Homeless Management Information System (HMIS). 

HMIS is both (1) a set of HUD data standards specifying what and how homelessness information should be collected, stored, shared, and reported, and (2) the shared database CoCs use to implement those standards. In this paper, we follow common usage among practitioners and use ``HMIS'' to mean the shared data system. HMIS collects standardized client-level data on enrollments in services such as emergency shelters, housing programs, and supportive services. With clients' consent, providers enter clients' service records into HMIS and in turn they gain visibility to clients' history within the network. This allows CoCs to generate un-duplicated client count and coordinate service delivery.

Our case study examines FL-507 Central Florida Continuum of Care, covering Orange, Osceola, and Seminole Counties, including Orlando metropolitan area. FL-507 includes 47 partner agencies, as well as a dedicated HMIS team and HMIS Advisory Committee overseeing implementation, training, data quality, and reporting. They also maintain a publicly available dashboard for their services \cite{hmisfl_dashboards_reports}.

\section{Research Method}
We conducted six one-hour semi-structured interviews with stakeholders who manage and use HMIS in FL-507. Interviews were conducted virtually between August and September 2025 following IRB approval. Participants were recruited via snowball sampling, starting with HMIS lead contacts and expanding through referrals \cite{goodman1961snowball, atkinson2001accessing}. Our sample included four staff from the coordinating lead agency and two from participating service providers. Quotes are attributed using broad role labels to protect identities: [L] for lead agency staff and [P] for service provider staff. The interviewees’ roles span key responsibilities in data management as well as planning and operations. 

We use a protocol structured around the four stages of the data value chain which traces how data move from production to action: collection (record or gather data), storage (maintain, secure, and enable access to records), analysis (process data into indicators, reports, or models), and usage (apply outputs to decisions, accountability, or service delivery) \cite{lofgren2020value, ruijer2023social}. We asked consistent core questions about data practice and collaboration, and tailored follow-ups to participants’ roles. One author led all interviews with at least one additional author present to probe and support consistency. Interview questions are provided in Appendix~\ref{appendix:interview_questions}. Recordings were transcribed, cleaned, and de-identified. We conducted an inductive thematic analysis \cite{Braun2022} using open coding, with each transcript double-coded by two researchers (including the interviewer), followed by bottom-up affinity diagramming.


\section{Results}
For each stage, we first describe the data work performed and the actors responsible, then identify key themes and challenges related to collaboration.

\subsection{Data Collection}
Data collection is led by service providers’ frontline staff: case managers enter records and program managers oversee data quality. The CoC lead agency routinely monitors the timeliness and completeness of the data through quarterly scorecards, and provide feedback to enforce consistency in data practices. Providers also collect additional information to meet their own operational needs which they typically store separately. 

Participants described data collection as seeking a balance between compliance with HUD standards, operational efficiency, and fairness in service delivery. Funding and reporting obligations drive most data practices, as reimbursement depends on accurate documentation [L2, L1], and divergent funder expectations --- such as emphasizing service transactions versus case notes --- shape priority [L2]. As such, CoC prefers limiting collection to essential information to avoid redundancy, incompleteness, and privacy risks. But such simplification can make it harder to capture the details needed to assess program performance or equity outcomes [L3]. While standardization supports process equity through consistent and transparent handling of client data, participants noted that strictly adhering to data rules limits opportunities to provide inputs and address structural disparities. As [L4] reflected, describing strict prioritization guidelines with additional factors considered only when flagged by outreach staff or eligibility rules, ``\textit{I do wanna see the outcomes too,}'' recognizing that equal procedures do not necessarily produce equitable results.

Lack of trust is a major challenge in collaborative data collection. When clients believe their information may be shared without consent or used against them, they may withhold data, complicating data collection efforts [L3]. Mistrust also arises between the CoC and other government entities, especially the criminal justice system and law enforcement. [L4] emphasized that ``\textit{we don't want HMIS to be used to the detriment of the participants and we don't want law enforcement looking for people in HMIS.}'' In discussing interactions involving clients exiting jail or encountering law enforcement, [L2] noted that ``\textit{it is a fine line}'' between directing clients to the appropriate resource and protecting them. 

Uneven organizational capacity further constrains participation and buy-in in community-level data collection. Although data governance is centralized, effective implementation depends on local engagement. Implementing new data standards requires time and resources, increasing workload for staff and limiting coordination [P1]. While some organizations can invest heavily in digital infrastructure, many, particularly smaller providers, prioritize direct service delivery over data documentation, collecting data primarily to meet reporting requirements [L2, P1]. Many do not fully grasp the broader implications or the importance of community-level data, focusing instead on their own organizational needs. As [L2] explained, the CoC must continually build buy-in, not only for reimbursement but because data collection ``\textit{shapes and informs and gives us the best picture of where folks are at.}''

Finally, legal constraints and power asymmetry define the boundaries of collaborative data collection. Strict HMIS privacy requirements restricts collaboration with outside partners such as law enforcement, public defenders, jails, and hospitals. These limits are further shaped by unequal legal and administrative capacity, as larger partners can dictate the terms of data-sharing agreements. As [L1] observed, ``\textit{Hospitals have a lot of legal capital, [while] CoCs have no legal capital,}'' creating asymmetries in negotiating data-sharing agreements.

\subsection{Data Storage}
While data storage is centralized through HMIS, there is ongoing tension between centralization and customization. 

Providers often collect additional data for program management or service improvement which they store locally, and attempt to share the data with HMIS. However, the CoC noted that adding non-standardized data with HMIS is not always beneficial despite the providers’ best intent. Attempts to integrate these local data frequently result in duplication, high missing rates, creating confusion and gaps in collective knowledge and can negatively impact the clients [L2]. Thus, customization beyond standard HMIS requires careful dialogue between the CoC and providers to preserve data integrity [L3].

Providers also maintain a parallel paper-based system to promote equity in service access. Limited client digital literacy makes paper documentation more inclusive: ``\textit{only one out of ten felt comfortable turning on a computer,}'' making tasks such as typing or completing web forms ``\textit{a huge barrier}'' [P1]. This dual system reflects a pragmatic accommodation to equity concerns, even as it complicates efforts to maintain a unified data infrastructure.

\subsection{Data Analysis}
Data analysis in the network combines technical tools with committee-based oversight and serves two primary purposes: compliance/accountability, and learning/improvement. The lead agency produces performance reports for CoC committee planning and for federal/state reporting, and shares findings with homelessness advocacy networks. Providers typically use standard built-in analysis tools in HMIS for internal monitoring and external reporting. Participants wanted analysis to support system-level learning and storytelling, but in practice it remains mostly compliance-driven.

Joint sensemaking around HMIS data is difficult because interpretation requires both technical skills and deep contextual knowledge of how homelessness systems operate. The same metric can support very different narratives: contextualizing low acceptance rates alongside full fund utilization provides the basis to advocate for additional resources rather than demonstrating failure [P1]. Without shared interpretive frameworks, even accurate data can lead to conflicting conclusions about trends and outcomes. Developing a shared appreciation for data nuances can strengthen collaboration by enabling more accurate, contextually grounded interpretations of homelessness realities [L1, L2].

Organizational capacity further limits who can meaningfully participate in collaborative data analysis. Many smaller organizations lack staff who can both analyze and interpret data, limiting actionable insights and increasing reliance on off-the-shelf HMIS reports and a few key individuals. As [L1] observed, ``\textit{it is very difficult to obtain actionable intelligence because we don’t have people who can see both the macro and the micro.}'' In contrast, larger or longer-established agencies have more influence over setting data standards and collection methods, and are better equipped to conduct deeper analyses [P2]. Larger organizations also tend to prefer using data as the primary basis for measurable change [P1], while smaller organizations use it mainly to validate experiential judgment [P2]. Thus, organizations prioritize data analysis differently, limiting shared investment in collaborative analysis.

Finally, joint analytics is hindered when organizations do not find mutual benefits and shared objectives [L2]. For example, agencies like Medicaid providers may focus on using access to clients to advance their own portfolios rather than support shared analytical goals [L1]. Disagreements over which performance measures matter reflect deeper differences in expertise and organizational incentives. As a result, aligning priorities, securing buy-in, and sustaining joint analysis remain difficult.

\subsection{Data Usage}
Data use in the mandated network centers on data communication and decision-making. The CoC and providers use community meetings and a committee-to-board process to develop proposals and make system decisions, and convene town halls to communicate to the public. Consistent data communication helped providers build trust with government partners and the broader community. Participants described becoming trusted local experts by providing hyper-local data [P1], while town halls evolved from contentious debates into more informed, solution-oriented discussions [P2].

At the same time, coordinated data communication remains challenging due to differing expertise and assumptions between providers and funders. As noted, ``\textit{curating data for stakeholders is one of the most difficult aspects because they come with their own set of assumptions}'' [L1]. Stakeholders often seek evidence of ``success'' without shared agreement on what success means or how it should be measured. Without careful framing, data can reinforce stereotypes or obscure the complexity of homelessness [L3, L2].

To mitigate this, participants expressed that effective communication relies on clarity, transparency, careful framing, and storytelling [L2, P1, P2]. Qualitative contexts and storytelling are important especially when key data are missing. For example, HMIS does not track identification status, leaving the scope of service barriers for clients without IDs unclear. Acknowledging such gaps can prompt discussion about if and how new data elements should be collected and inform future data practices [L2].

\section{Discussion}
Using a case study from U.S. homelessness services, we show that shared data infrastructures do not merely process data; they institutionalize governance logic that defines whose knowledge counts and whose participation is feasible. Standardization supports interoperability and reporting, but it limits local customization. Resource and expertise gaps constrain collaborative data practices, and participation remains uneven: larger providers with analytical capacity can use data strategically, while smaller providers primarily participate to meet compliance requirements. Across sectors, legal and privacy constraints, mistrust, and misaligned incentives add friction.

Advancing equitable collaboration requires reframing data governance from a compliance regime toward a learning infrastructure that connects standardization with participatory interpretation. Practically, this means expanding analytical capacity among smaller providers, embedding community voice in data standards, and designing feedback loops that circulate insights across governance levels. Networks can support local priorities through layered standards: a stable reporting core, sanctioned local extensions, and clear pathways for experimentation. To facilitate shared interpretation, networks can pair shared metric definitions with documentation of underlying assumptions and common sources of error, and separate data quality checks from narrative interpretation.

This study, however, focuses on a single CoC and may not capture variations across policy and governance contexts. Future research could employ comparative and longitudinal designs to examine how shared data infrastructure evolve over time and across institutional settings. Further inquiry might also explore how local experimentation in data governance reshapes the balance between collaboration and control.

\bibliographystyle{ACM-Reference-Format}
\bibliography{sample-base}

\appendix
\section{Semi-Structured Interview Questions}
\label{appendix:interview_questions}
The core interview questions were organized around the four stages of the data value chain—collection, storage, analysis, and usage—to ensure comparability across interviews while allowing flexibility for probing and follow-up. Questions in each stage addressed the what (types of data), how (methods and tools), who (responsible actors), and collaboration (cross-sector coordination), along with an open prompt about perceived barriers and strengths.

\begin{table*}[t]
\centering
\centering
\small
\begin{tabular}{p{4cm} p{11cm}}
\toprule
\textbf{Data Value Chain Stage} & \textbf{Core Interview Questions} \\
\midrule

Data Collection (Acquisition) 
  & Q1. What types of data are gathered by the CoC/your organization? \\
  & Q2. How is data collected? \\
  & Q3. Who collects it? Who participates in developing or revising your data collection tools (e.g., staff, people with lived experience, partner agencies)? \\
  & Q4. [Collaboration] How do you coordinate data collection with other organizations or members of the CoC (e.g., government agencies, nonprofits, community groups)? \\
  & Q5. [Open] Barriers and strengths in the data collection stage. \\
\midrule

Data Storage (Processing, Cleaning)
  & Q1. What systems or platforms does your organization use to store client and program data (e.g., HMIS, internal databases, cloud services, paper files)? \\
  & Q2. How do you decide what data storage solutions to use for different types of information? \\
  & Q3. Who is authorized to access stored data, and how are access permissions managed or monitored? \\
  & Q4. [Collaboration] How do you handle data storage when collaborating within or with government agencies compared to nonprofit organizations? Do you use shared systems or separate platforms? \\
  & Q5. [Open] Barriers and strengths in the data storage stage. \\
\midrule

Data Analysis
  & Q1. What types of analyses do you or the CoC/your organization perform regularly (e.g., population demographics, trend analysis, service usage, funding, performance evaluation, racial equity assessments)? \\
  & Q2. How is collected data used to identify trends and gaps in services? What analytic tools or software do you use, and are they sufficient for your needs? \\
  & Q3. Who is responsible for analyzing the data, a single agency or multiple agencies? Are there any capacity or training challenges? \\
  & Q4. [Collaboration] How does your approach to data analysis differ when working within or with government agencies compared to nonprofit organizations? \\
  & Q5. [Open] Barriers and strengths in the data analysis stage. \\
\midrule

Data Usage (Publication, Impact)
  & Q1. What types of data does your organization use to inform programs, decision-making, or advocacy efforts? What outcomes or changes have resulted from using data in your organization? \\
  & Q2. How do you share findings with stakeholders (e.g., public reports, internal meetings)? \\
  & Q3. Who is responsible for using and interpreting data in your organization? \\
  & Q4. [Collaboration] How do your organization and partners from other sectors (such as government, nonprofit, or the private sector) collaborate to use and interpret data? \\
  & Q5. [Open] Barriers and strengths in the data usage stage. \\
\bottomrule
\end{tabular}
\caption{Core interview questions by data value chain stage.}
\label{tab:data_value_chain_interview_simple}

\end{table*}

\end{document}